\documentclass[10pt,a4paper]{article}

\usepackage{times}

\usepackage{graphicx,subfigure,epsfig}
\usepackage{epsfig,wrapfig}
\usepackage{epstopdf} % for pdflatex

\usepackage{fancyhdr}

\usepackage[lmargin=2.5cm, rmargin=2.5cm,tmargin=3.50cm,bmargin=3.50cm]{geometry}
\usepackage{indentfirst}

\usepackage{titlesec}
\titleformat{\section}[hang]
  {\centering}{\thesection}{1ex}{\normalsize \textsc}%%
\titleformat{\subsection}[hang]
  {}{\thesubsection}{1ex}{\normalsize \textit}%%

\renewcommand{\thesection}{ \normalsize \textnormal{\Roman{section}.}}
\renewcommand{\thesubsection}{\normalsize \textnormal{\textsc{\textit{\Alph{subsection}.}}}}

\def\e{\begin{equation}}
\def\f{\end{equation}}
\def\_#1{{\bf #1}}

\def\.{\cdot}

\begin{document}

%%% Title of paper
\title{\large \textbf{Topological transition in a nanowire medium and its radiative implication}}
%
%%% Author(s) and affiliation
\def\affil#1{\begin{itemize} \item[] #1 \end{itemize}}
\author{\normalsize \bfseries \underline{M.S.~Mirmoosa$^1$}, S.Yu.~Kosulnikov$^{1,2}$ and C.R.~Simovski$^1$\\}
\date{}
\maketitle
\thispagestyle{fancy} % header also to the first page
\vspace{-6ex}
\affil{\begin{center}\normalsize $^1$Department of Radio Science and Engineering, School of Electrical Engineering, Aalto University, P.O.~Box 13000, FI-00076 AALTO, Finland\\
$^2$ITMO University, International Research Center of Metamaterials, 197043, St.~Petersburg, Russia\\
mohammad.mirmoosa@aalto.fi
 \end{center}}

%%% Abstract
\begin{abstract}
\noindent \normalsize
\textbf{\textit{Abstract} \ \ -- \ \
%%% Start here with text of abstract
We reveal and study the topological transition in a metamaterial formed by parallel nanowires of polaritonic material. When the dispersion 
transits from the elliptic (epsilon-positive) to hyperbolic (epsilon-indefinite) regime, a very specific isofrequency surface arises which 
implies an extraordinary Purcell factor in spite of noticeable optical losses.}
\end{abstract}

\section{Introduction}

Wire media -- an important class of hyperbolic metamaterials -- were reviewed in \cite{simovski}, where it was noticed that their key feature is strong spatial dispersion. The spatial dispersion distorts the isofrequency surfaces of regular wire media making them qualitatively different from hyperbolic ones -- either flattened or locally deviating their general shape in the vicinity of the $\Gamma$ point of reciprocal lattice. In the available literature, wire media are characterized by an indefinite complex permittivity tensor -- a uniaxial tensor whose components have either positive or negative sign of the real part, whereas the imaginary part of these components keeps reasonably small. Of course, both axial and transverse components of this tensor may be also negative, but it is not interesting since the wire medium (WM) becomes opaque. More interesting is to notice that both these components can be positive and the WM may become an anisotropic dielectric with elliptic type of dispersion.

Below we show that both hyperbolic and elliptic regimes are achievable with nanowires prepared of so-called polaritonic material, such as LiTaO3. In such WM, the transverse component of the effective permittivity tensor keeps positive whereas the axial component (its real part) changes the sign over the frequency axis in the mid IR range. This means that there is a frequency at which the hyperbolic dispersion regime transits to the elliptic one and vice versa. In accordance to \cite{krishnamurty}, this regime is called topological transition. Topological transition in a hyperbolic metamaterial implemented as a stack of metal-dielectric bilayers is already known \cite{yang}. On condition of low optical losses this regime is characterized by extraordinary density of photonic states \cite{krishnamurty,yang}. This density of states implies the so-called Purcell effect, and one of the most important implications of such a regime is a high Purcell factor -- enhancement of the decay rate of a quantum  emitter submerged into such a medium. Equivalently, this factor can be defined as the enhancement of the radiated power of a monochromatic subwavelength dipole source \cite{poddubny}.

In this paper, we show that the radiation of a short dipole is strikingly enhanced in spite of rather high optical losses in our WM which would have obviously suppressed the Purcell effect
in a usual hyperbolic medium. The reason of high Purcell factor is the combination of elliptic and the hyperbolic isofrequencies in the reciprocal space. The two branches are connected at two points where $\rm Re(\varepsilon_{\parallel})=0$. This becomes possible because the spatially dispersive effective permittivity depends on the axial component of the wave vector. As a result, neither at high nor at low spatial frequencies there is a cut-off: in the effective-medium approximation the spatial spectrum is unbounded. Another feature is the high Purcell factor for an axial dipole -- in recent studies published after \cite{poddubny} the radiation of an axial dipole is suppressed, and the Purcell effect holds only for a transversely polarized dipole.

\section{Theory}
\begin{figure}[t!]
\centering \epsfig{file=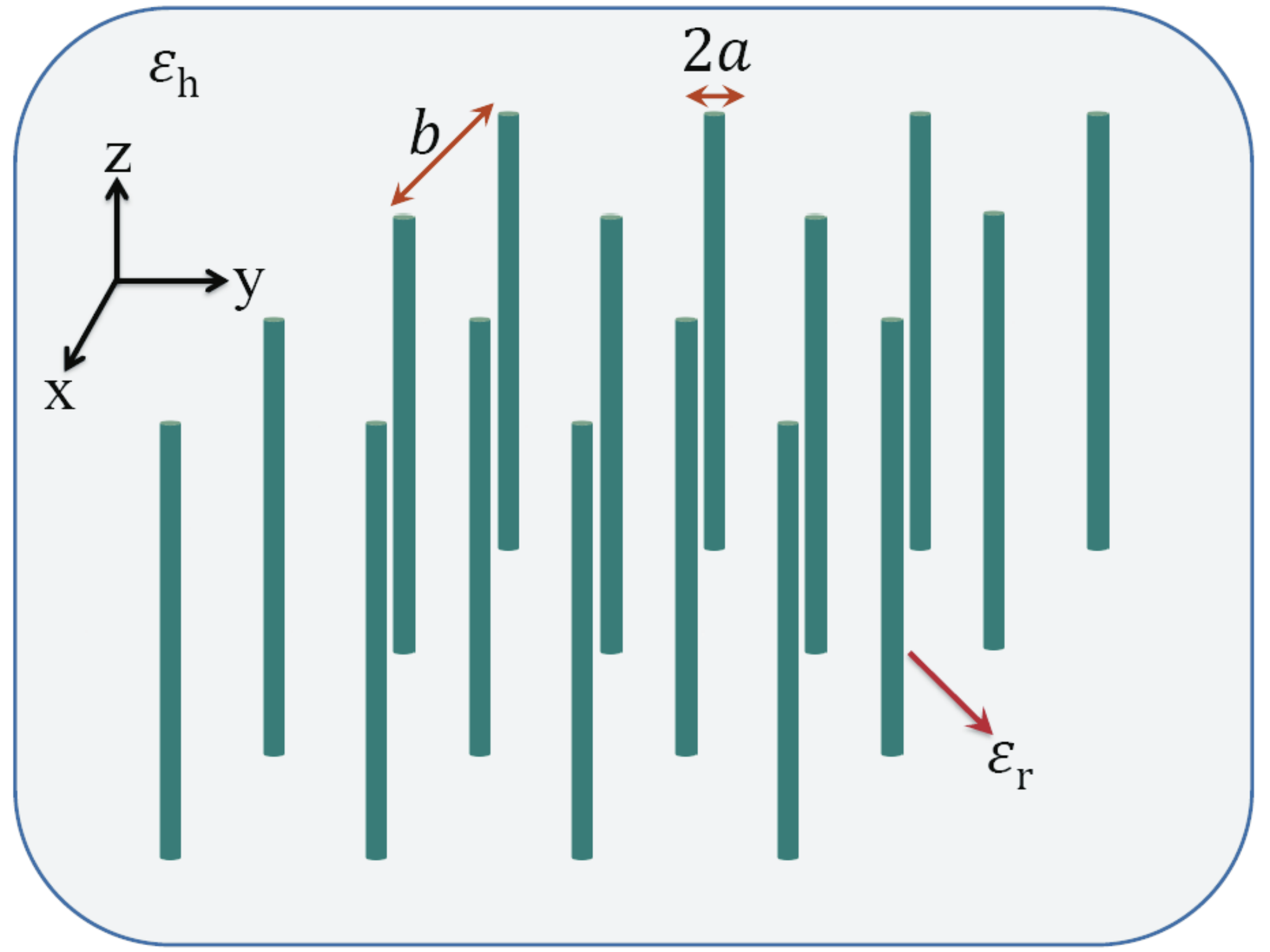, width=0.5\textwidth}
\caption{Long parallel wires ($\varepsilon_{\rm{r}}$) which are set in a square lattice. $b$ and $a$ are the lattice constant and the radius of a dielectric rod, respectively, and $\varepsilon_{\rm{h}}$ represents the host medium.}
\label{wire_medium}
\end{figure}

Optical properties of a WM sketched in Fig.~\ref{wire_medium} in the effective-medium model is described by a uniaxial dyad
\\
\begin{equation}
\overline{\overline{\epsilon}}=\varepsilon_\perp\mathbf{x_0}\mathbf{x_0}+\varepsilon_\perp\mathbf{y_0}\mathbf{y_0}+\varepsilon_\parallel\mathbf{z_0}\mathbf{z_0},
\end{equation}
where the transverse ($\varepsilon_{\perp}$) and axial ($\varepsilon_{\parallel}$) components, in case of thin wires performed of negative-epsilon material, are given by the well-known spatially-dispersive model by Silveirinha (see e.g. in \cite{simovski}):
\begin{equation}
\varepsilon_{\perp}=1+\displaystyle\frac{2}{\displaystyle\frac{\varepsilon_{\rm{r}}+1}{f_{\rm{v}}(\varepsilon_{\rm{r}}-1)}-1},\,\,\,\,\,\,\,\,\,\,\,\,\,\,\,
\varepsilon_{\parallel}=1+\displaystyle\frac{1}{\displaystyle\frac{1}{f_{\rm{v}}(\varepsilon_{\rm{r}}-1)}-\displaystyle\frac{k_0^2-\beta^2}{k_{\rm{p}}^2}}.
\label{eq:effper}
\end{equation}
Here, $\varepsilon_{\rm{r}}$ is the relative dielectric constant of the wires, $f_{\rm{v}}$ denotes the volume fraction, $k_0$ is the free-space wave number and $\beta$ represents the axial component of the wave vector. Also, $k_{\rm{p}}$ is called the plasma wave number expressed as $k_{\rm{p}}=\displaystyle({1}/{b})\sqrt{{2\pi}/[{0.5275+\ln{(\displaystyle{b}/{2\pi a})}}]}$ ($a$ is the radius of the wires and $b$ denotes the lattice constant). Equation~\ref{eq:effper} assumes that wires are embedded in free space. What is important in Eq.~\ref{eq:effper} is that the axial component of the permittivity depends on the component of the wave vector parallel to the rod wires. In other words, the wire medium is spatially dispersive.

In wire medium since it is a kind of uniaxial anisotropic medium, the transverse component of the wave vector is written as $q^2=\varepsilon_{\perp}k_0^2-\beta^2$ which corresponds to ordinary or TE-waves, and $q^2=\varepsilon_{\parallel}[k_0^2-\displaystyle{\beta^2}/{\varepsilon_{\perp}}]$ that corresponds to extraordinary or TM-waves. In this work, only TM-waves are important because in the vicinity of the short dipole, the reactive electromagnetic power stored as the evanescent waves are related to transverse magnetic (TM) waves. Based on the formula given above, if we assume that $\varepsilon_\perp$ is larger than zero, TM-waves can propagate within the medium if either 1) $k_0^2-\beta^2/\varepsilon_\perp>0$ and $\varepsilon_\parallel>0$, or 2) $k_0^2-\beta^2/\varepsilon_\perp<0$ and $\varepsilon_\parallel<0$. Both conditions can be implemented at a single frequency: the first case resulting in an elliptic dispersion corresponds to $\beta<k_0\sqrt{\varepsilon_\perp}$ and the second case giving the hyperbolic dispersion -- to $\beta >k_0\sqrt{\varepsilon_\perp}$.

WM grants both these possibilities since its axial component of the effective permittivity ($\varepsilon_\parallel$) depends on $\beta$. Assume that for $\beta=k_0\sqrt{\varepsilon_\perp}$ at a certain frequency, $\varepsilon_\parallel$ gets zero. After some algebraic manipulations based on Eq.~\ref{eq:effper}, we can conclude that $\varepsilon_{\rm{r}}=1+k^2_{\rm{p}}/[f_{\rm{v}}(k_0^2-\beta^2-k^2_{\rm{p}})]\approx1-1/f_{\rm{v}}<0$. Now, for $\beta>k_0\sqrt{\varepsilon_\perp}$, it can be shown that $\varepsilon_\parallel<0$, for $\beta<k_0\sqrt{\varepsilon_\perp}$, $\varepsilon_\parallel<0$, and at $\beta=k_0\sqrt{\varepsilon_\perp}$: $\varepsilon_\parallel=0$. In other words, in wire medium we can join hyperbolic and elliptic dispersions together at the point where $\varepsilon_\parallel=0$ and this is possible when the material of nanowires has negative permittivity $\varepsilon_{\rm{r}}$.

To confirm our expectation on the Purcell factor, we have done several full-wave simulations using the CST software. For example, the fraction ratio $f_{\rm{v}}=0.0804$ may correspond to $a=32$ nm and $b=200$ nm. Therefore, the value $\varepsilon_{\rm{r}}\approx-11.4$ is that allowing $\varepsilon_\parallel=0$ at a point $\beta=k_0\sqrt{\varepsilon_\perp}$. This value for $\varepsilon_{\rm{r}}$ corresponds to the permittivity of lithium tantalate (LiTaO3) at the frequency $f=39$ THz. Since the isofrequency contour in the plane $(q-\beta)$ is stretched along the axis $\beta$, the high Purcell factor should be observed for a dipole creating the spatial spectrum which broadens over the optical axis than in the transverse plane. This implies the axial polarization of the dipole. Though in work \cite{poddubny} the high Purcell effect in WM is claimed for a transverse dipole producing more TM-polarized spatial harmonics, for our purposes, the isofrequency shape turns out to be more important, and the Purcell effect is more pronounced for an axial dipole. 

\begin{figure}[t!]
\centering
\subfigure[]
{\includegraphics[width=6cm]{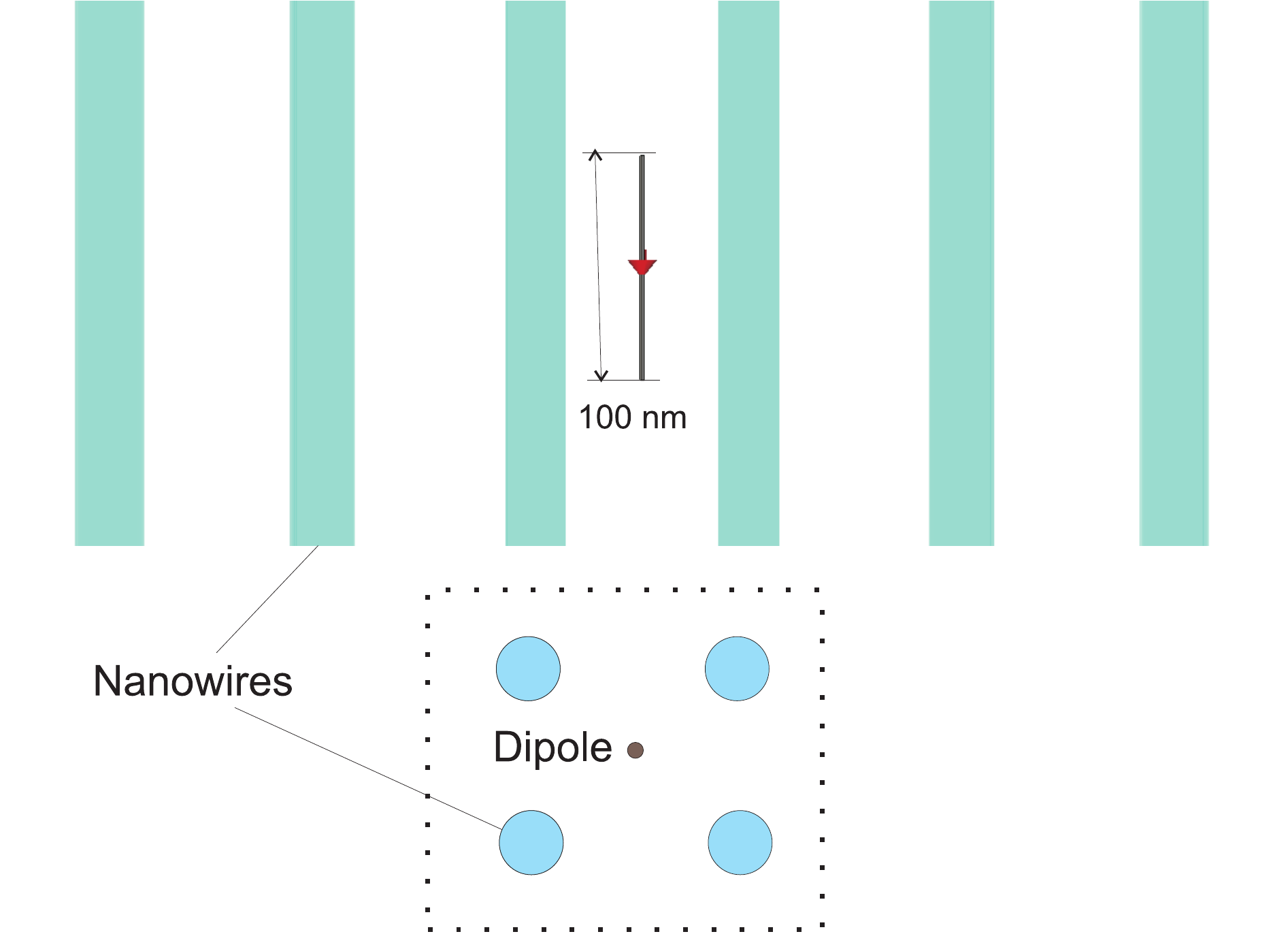}\label{fig:litao3_wires}}
\subfigure[]
{\includegraphics[width=7cm]{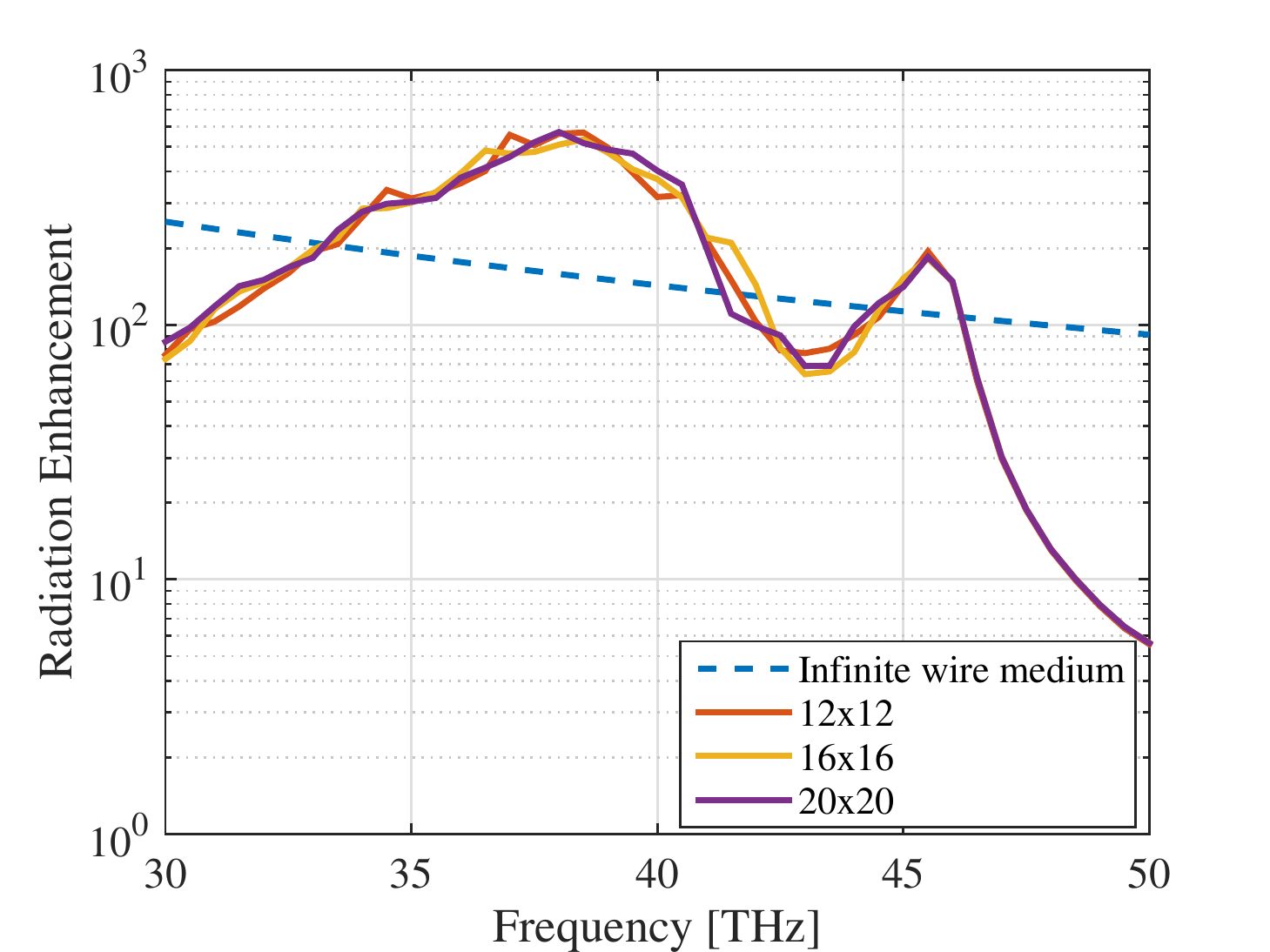}\label{fig:enhancement}}
\caption{(a)-- Dipole location and orientation in a finite-sized cubic sample of WM. (b)--Radiation enhancement of the dipole for three different sizes of the sample: 12$\times$12, 16$\times$16 and 20$\times$20 nanowires with period $b=$200 nm. The blue dashed curve corresponds to the radiation enhancement of a transversal dipole in an infinite PEC (perfect electric conductor) wire medium with same geometry.}
\label{fig:2}
\end{figure}

We simulated an electrically short dipole located at the center of a cubic sample of WM as it is shown in Fig.~\ref{fig:2}. At $f=38$ THz, the radiation intensity becomes enhanced 570 times compared to the absence of the WM. This enhancement was found as the radiation resistance of the dipole normalized to that calculated in free space. To get rid of dimensional resonances, we performed simulations for three different sizes of the sample: 12$\times$12, 16$\times$16 and 20$\times$20 nanowires increasing similarly the length of nanowires. The blue dashed curve corresponds to the analytically calculated enhancement of radiation for a transverse dipole in an infinite WM \cite{poddubny}. At the frequency of the topological transition, this value is 3.8 times lower than that for the axial dipole embedded in cubic sample of LiTaO3 WM. We numerically checked -- if our nanowires are PEC, the radiation enhancement vanishes for both axial and transversal dipoles. We also have done simulation for different values of the array period, wire radius, and different positions of the dipole in the cubic sample, which all confirmed our expectations.

\section{Conclusion}

In this work, we have pointed a topological transition in a medium of parallel nanowires from epsilon-negative material. Applying the spatially dispersive model we showed that
the axial component of the effective permittivity considered as a function of both frequency and axial component of the wave vector vanishes at one value of this dual argument. This transition grants both elliptic and hyperbolic dispersion and in spite of rather high optical losses results in the significant (2--3 orders of magnitude) increase of the radiated power for a subwavelength dipole submerged in such the medium. It is important that this huge Purcell factor was achieved for an axially oriented dipole. We have done extensive full-wave numerical simulations with finite samples of the WM which confirmed our theoretical expectations. Notice, that we also studied the radiation of the submerged dipole into free space (radiative Purcell factor) which is also enhanced by our WM (by one order of magnitude) and demonstrates a specific pattern. These results will be shown in our presentation.

%%% References

\end{document}